\documentclass[aps,prresearch,reprint,showpacs,groupedaddress,amsmath,amssymb,superscriptaddress,nofootinbib]{revtex4-1}

\usepackage{graphicx}
\usepackage{dcolumn}
\usepackage{bm}
\usepackage{color} 
\usepackage{siunitx}
\usepackage{pgf}
\usepackage[colorlinks=true,linkcolor=blue,citecolor=blue,urlcolor=blue, anchorcolor=red, filecolor=red, menucolor=red]{hyperref}
\usepackage[caption=false]{subfig}

\usepackage{multirow}

\begin{document}

\title{Efficiency and beam quality for positron acceleration in loaded plasma wakefields}

\author{C. S. Hue}
\thanks{These authors have contributed equally to this work.}
\affiliation{LOA, ENSTA Paris, CNRS, Ecole Polytechnique, Institut Polytechnique de Paris, 91762 Palaiseau, France}

\author{G. J. Cao}
\thanks{These authors have contributed equally to this work.}
\affiliation{LOA, ENSTA Paris, CNRS, Ecole Polytechnique, Institut Polytechnique de Paris, 91762 Palaiseau, France}
\affiliation{Department of Physics, University of Oslo, NO-0316 Oslo, Norway}

\author{I. A. Andriyash}
\affiliation{LOA, ENSTA Paris, CNRS, Ecole Polytechnique, Institut Polytechnique de Paris, 91762 Palaiseau, France}

\author{A. Knetsch}
\affiliation{LOA, ENSTA Paris, CNRS, Ecole Polytechnique, Institut Polytechnique de Paris, 91762 Palaiseau, France}

\author{M. J. Hogan}
\affiliation{SLAC National Accelerator Laboratory, Menlo Park, CA 94025, USA}

\author{E. Adli}
\affiliation{Department of Physics, University of Oslo, NO-0316 Oslo, Norway}

\author{S. Gessner}
\affiliation{SLAC National Accelerator Laboratory, Menlo Park, CA 94025, USA}

\author{S. Corde}
\email[Corresponding author: ]{sebastien.corde@polytechnique.edu}
\affiliation{LOA, ENSTA Paris, CNRS, Ecole Polytechnique, Institut Polytechnique de Paris, 91762 Palaiseau, France}

\date{\today}

\begin{abstract}
Accelerating particles to high energies in plasma wakefields is considered to be a promising technique with good energy efficiency and high gradient. While important progress has been made in plasma-based electron acceleration, positron acceleration in plasma has been scarcely studied and a fully self-consistent and optimal scenario has not yet been identified. For high energy physics applications where an electron-positron collider would be desired, the ability to accelerate positron{s} in plasma wakefields is however paramount. Here we show that {the preservation of beam quality can be compromised in a plasma wakefield loaded with a positron beam,  and a trade-off between energy efficiency and beam quality needs to be found.} For electron beams driving linear plasma wakefields, we have found that despite the transversely nonlinear focusing force induced by positron beam loading, the bunch quickly evolves toward an equilibrium distribution with limited emittance growth. Particle-in-cell simulations show that for \si{\micro\metre}-scale normalized emittance, the growth of uncorrelated energy spread sets an important limit. Our results {demonstrate} that the linear or moderately nonlinear regimes with Gaussian drivers provide a {good} trade-off, achieving simultaneously energy-transfer efficiencies exceeding \SI{30}{\%} and uncorrelated energy spread below \SI{1}{\%}, while donut-shaped drivers in the nonlinear regime are more appropriate to accelerate {high-charge} bunches at higher gradients, at the cost of a degraded trade-off between efficiency and beam quality.
\end{abstract}

\maketitle

\section{Introduction}
\label{sec:intro}

Particle accelerators based on radio-frequency technology are being used in a very broad range of applications, from free-electron lasers or medicine to particle colliders for high-energy physics. For the latter, the limited accelerating gradient of this technology {renders the footprint and the cost for future machines prohibitively expensive}. Plasma-based acceleration, driven by particle beams~\cite{Veksler1956, Fainberg1960, Fainberg1968, Chen1985} or laser pulses~\cite{Tajima1979, Esarey2009}, is a potential candidate to considerably increase the accelerating gradient and to provide efficient high-energy particle accelerator{s}. In the last decades, substantial progress has been made in electron acceleration using plasma-based accelerators in the nonlinear bubble or blowout regime~\cite{Rosenzweig1991, Pukhov2002, Malka2002, Faure2004, Geddes2004, Mangles2004, Lu2006a, Blumenfeld2007}, a regime that is particularly well suited for high energy efficiency~\cite{Tzoufras2008, Litos2014, Lindstrom2021} and whose field structure is ideal for beam quality preservation~\cite{Clayton2016}. 
However, for applications {towards high-energy colliders, it is imperative for plasma-based accelerators to be capable of accelerating positrons. Because of the need for high luminosity}, positron beams must also be accelerated with high energy efficiency and have very small final normalized emittances, on the order of {$10$ to $\SI{100}{nm}$~\cite{ilc, clic}}. Moreover, the positron beam energy spread needs to be kept below {percent} level so that the beam can be focused to the interaction point by a final focus system~\cite{Raimondi2001}. This requires keeping both the correlated energy spread (different longitudinal slices having different energies) and the uncorrelated (or slice) energy spread under control. Finally, the acceleration process needs to be stable and to reach full depletion of the driver. 

The above criteria, namely energy efficiency, emittance, energy spread and stability, must be {optimized for simultaneously in a high-energy collider application}. This is particularly challenging in the case of plasma-based positron acceleration. Experimentally, positron acceleration in plasma has been successfully demonstrated for positrons at the rear of a single drive bunch~\cite{Blue2003} with high energy efficiency~\cite{Corde2015}, as well as for distinct trailing positron bunches in both linear and nonlinear regimes~\cite{Doche2017}, but the question of beam quality and emittance preservation was not addressed. Positron bunches were also accelerated  in hollow plasma channels~\cite{Gessner2016a, Gessner2016b}, however, the stability of the hollow plasma accelerator was shown to be compromised by the presence of strong transverse wakefields~\cite{Lindstrom2018} that lead to the beam breakup instability~\cite{Schroeder1999}.

Novel methods, such as the use of {finite-radius} plasma columns~\cite{Diederichs2019, Diederichs2020} or donut-shaped electron~\cite{Jain2015} and laser~\cite{Vieira2014} drivers, have also been proposed to excite plasma wakefields with an extended region that can be focusing and accelerating for positrons, usually obtained by the creation of a long plasma electron filament in the vicinity of the propagation axis. Apart from the hollow plasma channel accelerator, methods to accelerate positrons in plasma typically have in common the presence of an excess of plasma electrons within the accelerated positron bunch. Such plasma wakefields can then be used to accelerate low-charge bunches while preserving their quality, for example with weak beam loading in the linear regime~\cite{Brinkmann2017}, but the energy efficiency with weak beam loading is too low to be of interest for a collider application. When increasing positron beam charge and energy efficiency, positron beam loading becomes the key challenge because of the quick response { of the plasma electrons to the positron load}. This quick response can lead to a transversely nonlinear focusing force {that can potentially drive emittance growth~\cite{Muggli2008}, as well as a transversely non-uniform accelerating field that can induce growth in} uncorrelated energy spread. The problem of positron beam loading in the presence of plasma electrons within the bunch has similarities to the physics of ion motion for electron acceleration in the blowout regime~\cite{Rosenzweig2005, Gholizadeh2010, An2017, Benedetti2017, Mehrling2018}, but is considerably more challenging, {as plasma electrons are much more mobile than ions}. A first insight into the physics of positron beam loading was reported in the nonlinear regime~\cite{Corde2015}, where beam loading allows the flattening of $E_z(\xi)$ {and induces a filament of plasma electrons, thereby reaching high energy efficiencies and providing focusing to the accelerated positrons}. There are also detailed studies in the specific context of the {finite-radius} plasma column~\cite{Diederichs2020}, and in the case of nonlinear and asymmetric plasma wakefields in hollow channels~\cite{Zhou2020}, where plasma electrons from the channel wall {cross the axis to} provide focusing to the positron beam.

In this article, we focus on plasma wakefields driven by Gaussian or donut-shaped electron beams in uniform plasmas, from the linear to the nonlinear regime. Positron beam loading is shown to induce a trade-off between energy efficiency and beam quality, and that for \si{\micro\metre}-scale normalized emittance, an important limit arises from the growth of uncorrelated energy spread driven by the transversely nonuniform accelerating field associated {with} the strong positron load. In Sec. \ref{sec:linear_wakefields}, we illustrate the positron beam loading problem for linear plasma wakefields. Section \ref{sec:energy_efficiency} presents an analytical model and simulation results for the {energy-transfer} efficiency in three dimensions, and shows that a mismatch in transverse size of the wakefield from the electron driver and the wakefield from the positron bunch can considerably reduce the energy efficiency. The latter suggests that higher efficiencies can be obtained with linear plasma wakefields for beam sizes smaller than the plasma skin depth. In Sec. \ref{sec:transverse}, the evolution of the transverse phase space of the positron bunch is discussed, and it is shown that when starting from quasi-matched conditions, the beam evolves rapidly toward an equilibrium, with limited emittance growth during this initial evolution, and emittance preservation afterwards. Section \ref{sec:longitudinal} discusses the evolution of the longitudinal phase space of the positron bunch, and shows that the growth of slice energy spread can pose serious limitations. In Sec. \ref{sec:donut}, the blowout regime with a donut-shaped electron driver is considered, where the donut allows for an excess of plasma electrons to be present near the propagation axis in the blowout cavity, {thus providing} focusing to the positron bunch. The properties of the plasma wakefields and the induced slice energy spread for the accelerated positron beam are presented and discussed. Section \ref{sec:espread_efficiency} presents a comparison of different regimes, considering similar initial positron beam parameters but optimizing the driver separately for each specific regime (see Sec.~\ref{sec:optimization}), so that their performance can be determined in terms of energy efficiency and uncorrelated energy spread, {illustrating} the trade-off between these two quantities (see Sec.~\ref{sec:tradeoff}). Conclusions are finally presented in Sec.~\ref{sec:conclu}.

\section{Linear plasma wakefields}
\label{sec:linear_wakefields}

\subsection{Energy-transfer efficiency}
\label{sec:energy_efficiency}

The {energy-transfer} efficiency $\eta$ is defined as the ratio of energy gained by the trailing bunch $W_{t,\mathrm{gain}}$ and energy lost by the driver $W_{d,\mathrm{loss}}$,
\begin{equation}\label{eq: eta definition}
    \eta = \frac{W_{t,{\text{gain}}}}{W_{d,{\text{loss}}}}=-\frac{Q_t\langle E_z\rangle_t}{Q_d \langle E_z\rangle_d},
\end{equation}
with $\langle E_z\rangle_{t,d}$ the longitudinal electric field averaged over particles in the bunch and over the propagation distance in the plasma, and $Q_{t,d}$ the bunch charge.
It represents the efficiency in the transfer of energy from the drive to the trailing bunch through the plasma, and can also be thought as the {energy-extraction} efficiency from the plasma to the trailing bunch ($W_{d,\mathrm{loss}}$ being the energy given to the plasma by the drive bunch). Importantly, this {energy-transfer} efficiency $\eta$ is a figure of merit of the wakefield and does not take into account the ratio between $W_{d,\mathrm{loss}}$ and the total initial energy in the drive bunch, which would represent the efficiency from drive to plasma and would depend on the acceleration distance that can be achieved. In the linear regime, beam loading can be understood from the fact that the total wakefield is simply the superposition of drive and trailing plasma wakefields, and in one dimension one finds for the {energy-transfer} efficiency~\cite{Katsouleas1987}:
\begin{equation}\label{eq: 1D model}
    \eta = \frac{N_t}{N_d}\left(2-\frac{N_t}{N_d}\right),
\end{equation}
where $N_{t,d}$ is the number of particles in each bunch, and we have assumed that the drive and trailing particle charges satisfy $|q_d|=|q_t|$, the drive-trailing bunch separation is { $\Delta\xi = \lambda_p/2 \pmod{\lambda_p}$ for $q_d=q_t$ [respectively $\Delta\xi = \lambda_p \pmod{\lambda_p}$ for $q_d=-q_t$]}, and either $\sigma_{dz}=\sigma_{tz}$ (same drive and trailing bunch lengths) or both bunches are short: $k_p\sigma_z\ll1$, with $k_p=\omega_p/c$ and $\omega_p=\sqrt{n_0e^2/(m_e\epsilon_0)}$ the plasma frequency for a plasma of density $n_0$.
{The parabolic relationship in Eq.~\eqref{eq: 1D model}} shows that the efficiency is {maximized} and equal to 1 when $N_t=N_d$, that is in the case of perfectly-destructive interference between drive and trailing wakefields corresponding to a vanishing plasma wave behind the trailing bunch, when all the energy in the plasma is extracted by the trailing bunch.

In three dimensions, the {energy-transfer} efficiency not only depends on $N_{t,d}$, but also on the size and shape of drive and trailing plasma wakefields. In particular, one cannot expect to reach perfectly-destructive interference ($\eta=1$) if the size and shape of drive and trailing wakefields differ. In general, $\eta$ will also depend on drive and trailing beam sizes ($\sigma_{dr}$ and $\sigma_{tr}$), bunch lengths ($\sigma_{dz}$ and $\sigma_{tz}$) and on the plasma skin depth $1/k_p$. In the linear regime, analytical calculations can be performed for separable bunch shapes, as shown in Appendix~\ref{Appendix_eta3D} and Eq.~(\ref{eq: theoretical model for efficiency general}), and can be simplified for large ($k_p\sigma_r\gg1$) Gaussian bunches to
\begin{equation}
\label{eq: efficiency for same dw bunch length}
    \eta = \frac{N_{t}}{N_{d}}\frac{\sigma_{dr}^2}{\sigma_{tr}^2}\left[\frac{4}{1+\frac{\sigma_{dr}^2}{\sigma_{tr}^2}}-\frac{N_t}{N_d}\right],
\end{equation}
under the same assumptions as Eq. (\ref{eq: 1D model}), and also takes the form of a parabolic relationship between $\eta$ and $N_t/N_d$, with parameters now depending on beam sizes.

\begin{figure}[t!]
\includegraphics[width=0.47\textwidth]{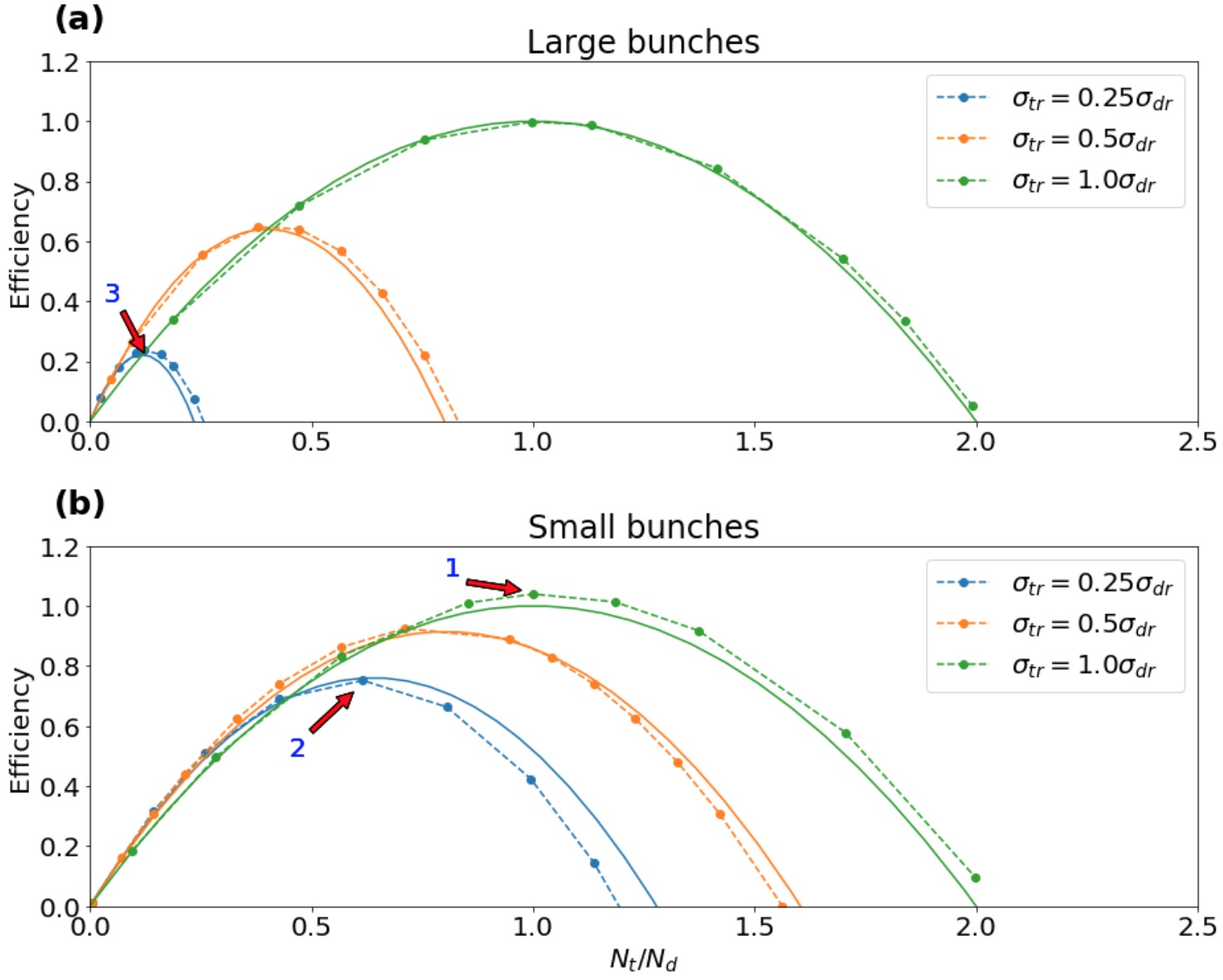}
\caption{\label{fig: Efficiency curves} Analytical (solid lines) and simulation (circles and dashed lines) results for energy-transfer efficiencies in 3D linear regime, for large beams $\sigma_r\gg 1/k_p$ (a) and small beams $\sigma_r < 1/k_p$ (b). The plasma density is set to $n_0=\SI{4e17}{\per\cubic\centi\metre}$ and the peak drive bunch density is $n_b=0.042\:n_0$ for all simulations. The drive bunch has $\sigma_{dr}=\SI{100}{\micro\metre}$ and $\sigma_{dz}=\SI{4}{\micro\metre}$ in (a) and $\sigma_{dr}=\sigma_{dz}=\SI{2}{\micro\metre}$ in (b). The trailing bunch has a beam size $\sigma_{tr}=0.25\:\sigma_{dr}$ (blue), $\sigma_{tr}=0.5\:\sigma_{dr}$ (orange) or $\sigma_{tr}=\sigma_{dr}$ (green), a bunch length $\sigma_{tz}=\sigma_{dz}$, and its charge is varied according to $N_t/N_d$. The simulated wakefields corresponding to the annotated points 1-3 are shown in Fig.~\ref{fig: Efficiency fields}.}
\end{figure}

\begin{figure}[t!]
\includegraphics[width=0.48\textwidth]{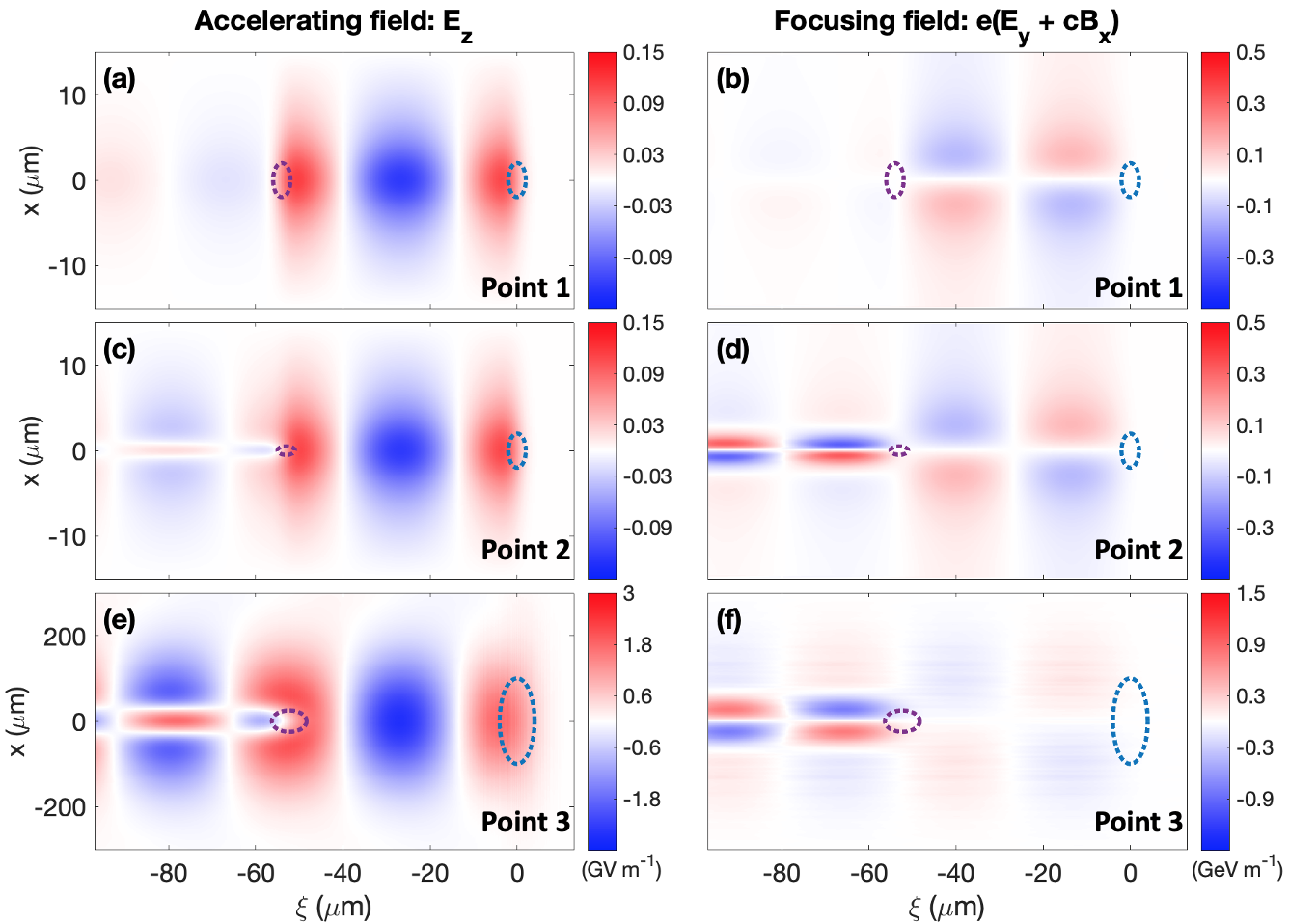}
\caption{\label{fig: Efficiency fields} Simulated wakefields [$E_z$ left column, $e(E_y+cB_x)$ right column] for the annotated points 1-3 in Fig.~\ref{fig: Efficiency curves}. Point 1 [(a)-(b)] corresponds to the small beam case with near-100\% efficiency ($\sigma_{tr}=\sigma_{dr}$), point 2 [(c)-(d)] to the small beam case with maximum efficiency for $\sigma_{tr}=0.25\:\sigma_{dr}$, and point 3 [(e)-(f)] to the large beam case with maximum efficiency for $\sigma_{tr}=0.25\:\sigma_{dr}$. Dashed ellipses show the location of drive and trailing bunches {with the $1\sigma$ contours of their bunch densities}.}
\end{figure}

In Fig.~\ref{fig: Efficiency curves}, the analytical results of Eqs.~(\ref{eq: efficiency for same dw bunch length}) and (\ref{eq: theoretical model for efficiency general}) are compared to numerical simulations performed using the Open Source quasi-static Particle-In-Cell (PIC) code QuickPIC~\cite{Huang2006, An2013, QuickPIC-OpenSource}. The plasma wakefield generated by both bunches is obtained from a single-step quasi-static simulation (no beam evolution), and the {energy-transfer} efficiency is computed from the simulated wakefield by averaging $E_z$ over each bunch [see Eq.~(\ref{eq: eta definition})], showing good agreement with the analytical results. For large beams, $k_p\sigma_r\gg1$, the maximum efficiency [see Eq.~(\ref{eq:eta_max})] and the corresponding value of the trailing charge [see Eq.~(\ref{eq:x_max})] both rapidly decrease when the trailing beam size is reduced with respect to that of the driver [see Fig.~\ref{fig: Efficiency curves}(a)]. This low efficiency can be understood by the strong mismach between the size of the wakefields of the drive and trailing bunches, as can be seen in Fig.~\ref{fig: Efficiency fields}(e). Indeed, in such a situation, the trailing wakefield can only overlap and cancel the drive wakefield over a small region of the drive wakefield, leaving a large amount of energy in the plasma wave {behind} the trailing bunch, thus leading to low efficiency, with $\eta_\mathrm{max}\simeq\SI{23}{\%}$ for $\sigma_{tr}=0.25\:\sigma_{dr}$ in Fig.~\ref{fig: Efficiency curves}(a). In contrast to large beams, for which the size of the wakefield is determined by the size of the beam, the wakefield for small beams typically extends over a plasma skin depth. As a result, a better overlap between drive and trailing wakefields is found for $\sigma_{tr}\neq\sigma_{dr}$ in {the small-beam case}, leading to higher maximum efficiencies corresponding to higher trailing charge, as seen in Fig.~\ref{fig: Efficiency curves}(b), with $\eta_\mathrm{max}\simeq\SI{74}{\%}$ for $\sigma_{tr}=0.25\:\sigma_{dr}$.

Figure~\ref{fig: Efficiency fields} shows simulated wakefields (longitudinal and transverse fields) for a few cases of interest. For $\sigma_{tr}=\sigma_{dr}$, the drive and trailing wakefields have exactly the same extent and shape, and it is therefore possible to approach $\eta=1$ with near-cancellation of the wakefield behind the trailing bunch [see Fig.~\ref{fig: Efficiency fields}(a)-(b)]. For the case { of small beams and $\sigma_{tr}=0.25\:\sigma_{dr}$, the wakefield can still be significantly weakened by the trailing bunch at transverse positions $|x|\gg\sigma_{tr}$ despite different beam sizes and different wakefield shapes for the drive and trailing bunches. This is because the fields extend typically over a distance $1/k_p\gg\sigma_{tr}$ [see Fig.~\ref{fig: Efficiency fields}(c)-(d)], thereby ensuring an efficient beam loading and high energy-transfer efficiencies.} {For beams with $k_p\sigma_r\gg1$, the transverse size of the wakefield is determined by the transverse beam size, and beam loading effects are localized within the trailing beam cross section, which limits the energy-transfer efficiency} [see Fig.~\ref{fig: Efficiency fields}(e)-(f)].

As it is generally desired to accelerate low emittance positron beams, especially for high energy physics applications, and given that the matched beam size for the trailing bunch is very small at low emittance and high energy (see Sec.~\ref{sec:transverse}), one can expect to have $\sigma_{tr}\ll\sigma_{dr}$. In this case, given the results presented above, the best strategy for reaching high {energy-transfer} efficiencies in the linear regime is therefore to consider small beams, and thus to work with a drive bunch with $k_p\sigma_{dr}\lesssim1$. In addition, while a linear wakefield can be driven by the drive bunch and used for positron acceleration, a matched positron bunch at low emittance and high energy can have a density easily exceeding that of the plasma, resulting in nonlinear beam loading (wakefield superposition no longer holds), which is nonlocal and acts on plasma electrons at distances much larger than $\sigma_{tr}$, which is favorable for high efficiencies.

\subsection{Transverse phase space and equilibrium}
\label{sec:transverse}

\begin{figure}[t!]
\includegraphics[width=0.54\textwidth]{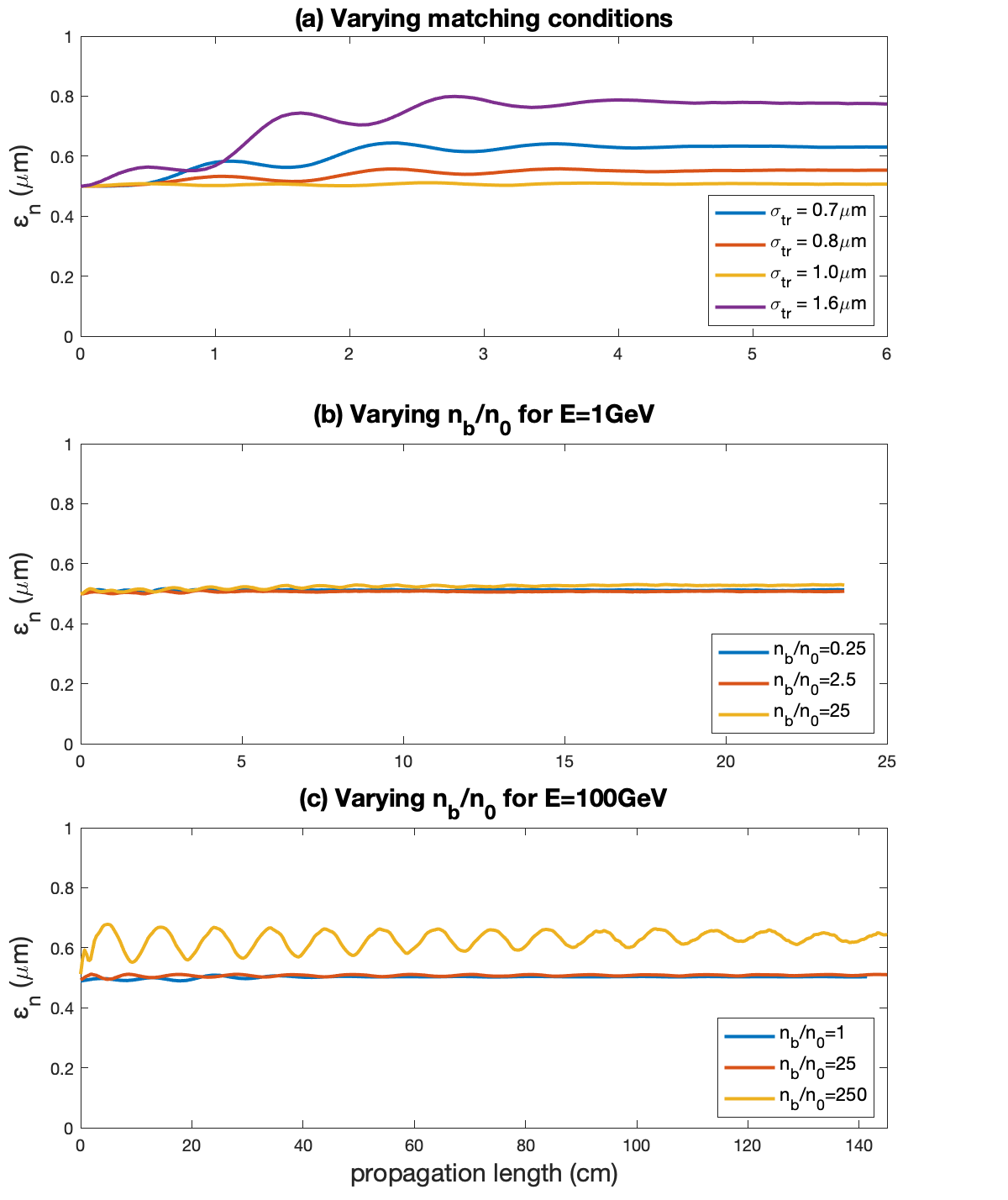}
\caption{\label{fig: emittance_evolution} Evolution of the normalized transverse emittance of the trailing positron bunch, $\varepsilon_n = \sqrt{\langle x^2\rangle \langle p_x^2\rangle-\langle x p_x \rangle^2}/mc$, during its propagation in the plasma. For all simulations, the plasma density is set to $n_0=\SI{5e16}{\per\cubic\centi\metre}$, the peak density of the electron drive bunch is $n_b=0.35\:n_0$, the drive beam size and bunch length are $\sigma_{dr}=\SI{14.5}{\micro\metre}$ and $\sigma_{dz}=\SI{16.7}{\micro\metre}$, the drive beam evolution is turned off and the drive to trailing bunch separation is $\Delta \xi=\SI{143}{\mu\metre}$. The initial trailing positron bunch energy is \SI{1}{GeV} in (a)-(b) and \SI{100}{GeV} in (c). The matching condition is varied in (a), while the trailing positron beam is quasi-matched for increasing values of $n_b/n_0$ in (b)-(c). All initial trailing positron beam parameters and corresponding values for efficiency and emittance growth are listed in Table~\ref{table: emittance growth study}.}
\end{figure}

While high energy efficiencies are desirable, it must be achieved with high quality positron beams and the acceleration process must preserve this quality. For the prospect of a plasma-based high-energy collider with a luminosity exceeding \SI{1e34}{cm^{-2}s^{-1}}, a normalized emittance on the order of \SI{10}{} to \SI{100}{nm} (in one transverse direction) is generally targeted, which represents a considerable challenge for the acceleration of positron beams in plasmas. Emittance preservation is possible in a transversely linear focusing force, which is the case for the acceleration of a matched electron beam in the blowout regime in the absence of ion motion~\cite{Clayton2016}. However, a transversely nonlinear focusing force, for example in the case of ion motion for electron acceleration in the blowout regime or plasma electron suck-in for positron acceleration, can induce emittance growth~\cite{An2017, Muggli2008}. In general, a non-Gaussian transverse equilibrium can be quickly reached by the beam, as is the case for a single decelerated beam in a linear wakefield~\cite{Lotov2017}, an electron beam in the blowout regime with ion motion~\cite{Benedetti2017, An2017}, or for a step-like focusing force~\cite{Diederichs2019}. While the equilibrium for a single decelerated beam has unusual and unwanted properties due to the absence of an external focusing field from a driver, such as on-axis singularity and very large transverse tails~\cite{Lotov2017}, the case of a trailing bunch in the focusing force from a driver has much more favorable equilibrium distributions, as shown in Ref.~\cite{Benedetti2017} for ion motion and Ref.~\cite{Diederichs2019} for the step-like focusing force of the finite-radius plasma column scheme. In particular, the emittance growth from an initially Gaussian-shaped trailing bunch to the equilibrium can be limited to about 10 to \SI{20}{\%} for ion motion~\cite{Benedetti2017, An2017}, and to approximately \SI{6}{\%} for the step-like focusing force~\cite{Diederichs2019}, and this emittance growth can be further reduced or eliminated by head-to-tail bunch shaping~\cite{Benedetti2017}. Although perfect emittance preservation is not possible for an initially Gaussian trailing bunch, the minimum emittance growth is obtained for a specific value of the beam size, which is referred to as the quasi-matched beam size.

\begin{table*}[t]
\begin{center}{}
\resizebox{0.8\textwidth}{!}{
\begin{tabular}{l|p{1.5cm}p{1.5cm}p{1.5cm}p{1.5cm}p{1.5cm}p{1.5cm}p{1.5cm}|p{1.5cm}p{1.5cm}}
& $\sigma_{tr}\:(\si{\micro\metre})$ & $\varepsilon_n\:(\si{\micro\metre})$ & $\beta\:(\si{\centi\metre})$ & $\sigma_{tz}\:(\si{\micro\metre})$ & $n_b/n_0$ & $k_b\sigma_{tz}$ & $E\:(\si{\giga\electronvolt})$ & $\eta$(\%) & $\Delta\varepsilon_n(\%)$\\
\hline
\multirow{4}{5em}{Fig.~\ref{fig: emittance_evolution}(a)}    & 0.7  & 0.5  &0.20  & 2.14  & 1 & 0.09 & 1 & 0.30 & 27.61\\
    &0.8    & 0.5   &0.26   & 2.14   & 1 & 0.09 & 1 & 0.39 & 11.58\\
    &1.0    & 0.5   &0.40   & 2.14   & 1 & 0.09 & 1 & 0.61 & 1.74\\
    &1.6    & 0.5   &1.02   & 2.14   & 1 & 0.09 & 1 & 1.55 & 55.40\\
\hline
\multirow{3}{5em}{Fig.~\ref{fig: emittance_evolution}(b)} & 1.01 & 0.5 & 0.41   & 2.14  & 0.25  & 0.045 & 1 & 0.16 & 1.74\\
& 1.00  & 0.5   &0.40   & 2.14  & 2.5   & 0.14 & 1 & 1.52 & 2.64\\
& 0.80  & 0.5   &0.26   & 2.14  & 25    & 0.45 & 1 & 9.15 & 5.83\\
\hline
\multirow{3}{5em}{Fig.~\ref{fig: emittance_evolution}(c)} & 0.327    & 0.5    &4.28   & 2.14      & 1  & 0.09 & 100 & 0.07 &2.73\\
& 0.288    & 0.5   &3.33   & 2.14      & 25  & 0.45 & 100 & 1.63 & 3.67\\
& 0.189    & 0.5   &1.43   & 2.14      & 250  & 1.4 & 100 & 5.24 & 30.01\\
\end{tabular}
}
\caption{Initial trailing positron beam parameters and corresponding values of efficiency and emittance growth for simulations presented in Fig.~\ref{fig: emittance_evolution}.}
\label{table: emittance growth study}
\end{center}
\end{table*}

Here, we study, by means of PIC simulations, the emittance evolution of an initially Gaussian trailing positron bunch in a quasilinear wakefield from the driver, which exhibits a transversely nonlinear focusing force. As we are interested in reaching high energy efficiencies, the trailing positron bunch can strongly load the wakefield and thus substantially modify the focusing force, and this positron beam loading can be nonlinear if $n_b/n_0>1$ for the trailing bunch. 
{ To model the potential emittance growth solely driven by the non-ideal properties of the wakefield and most importantly by positron beam loading, the driver evolution is turned off in the PIC simulations. By doing so, we do not consider other potential limits such as driver and wakefield stability over long propagation distances, and focus on the study of the evolution of the trailing bunch towards an equilibrium.} Figure~\ref{fig: emittance_evolution} shows the results of {QuickPIC-OpenSource} simulations performed for the same drive beam that excites a quasilinear wakefield, varying the initial parameters of the trailing positron bunch. The initial emittance of the trailing positron bunch is kept the same to \SI{0.5}{\micro\metre} in all simulations. Quasi-matching is, as expected, critically important, as can be observed in Fig.~\ref{fig: emittance_evolution}(a) for a 1 GeV initial trailing-bunch energy, where the emittance evolution is shown for different initial trailing beam sizes $\sigma_{tr}$, that is for different initial beta functions. Because of the nonlinear focusing force, a mismatched beam undergoes substantial emittance growth and quickly saturates after a few betatron periods, while the quasi-matched beam, with $\sigma_{tr} = \SI{1.0}{\micro\metre}$ and $\beta = \SI{0.4}{\centi\metre}$, reaches the equilibrium state with only negligible emittance growth.

Figure~\ref{fig: emittance_evolution}(b) illustrates the emittance evolution for different initial peak densities $n_b/n_0$ of the trailing positron bunch, by increasing its charge. Each simulation is optimized to determine the quasi-matched beam size and minimize emittance growth. Interestingly, the results show that the emittance growth of a quasi-matched beam does not substantially increase when $n_b/n_0$ becomes larger than one (comparing $n_b/n_0=0.25$ and $n_b/n_0=2.5$ with emittance growth of about \SI{2}{\%}), that is when beam loading becomes nonlinear. To test much higher values of $n_b/n_0$, increasing further the trailing charge completely overloads the wakefield (with part of the trailing bunch being decelerated), however, higher values of $n_b/n_0$ can be reached by increasing the trailing beam energy, which reduces the quasi-matched beam size and increases $n_b/n_0$. In Fig.~\ref{fig: emittance_evolution}(c), the same emittance evolution for increasing values of $n_b/n_0$ is shown but at a much higher initial trailing bunch energy of \SI{100}{\giga\electronvolt}, instead of \SI{1}{\giga\electronvolt} previously, corresponding to smaller quasi-matched beam sizes for similar trailing charges and energy efficiencies. Expressed in number of betatron periods, the propagation distance required to reach the equilibrium is larger at higher values of $n_b/n_0$, as well as the emittance growth, which is \SI{30}{\%} for the largest simulated trailing bunch density, $n_b/n_0=250$. Here, a transition in the characteristic regime of the plasma electron response to the positron beam occurs when $k_b\sigma_{tz}$ becomes larger than one, where $k_b$ is the plasma wave number associated with the positron beam density [{$k_b=\sqrt{n_be^2/(m_e\epsilon_0)}/c$}]. This transition was discussed in the case of a trailing electron beam~\cite{Lehe2014}, in which case a blowout is formed within the bunch itself when $k_b\sigma_{tz}\gtrsim1$. In the case of a trailing positron beam, plasma electrons are initially sucked in and will cross the propagation axis within the bunch itself if $k_b\sigma_{tz}\gtrsim1$, or even execute multiple oscillations if $k_b\sigma_{tz}\gg1$. In contrast, for $k_b\sigma_{tz}\ll1$, positron beam loading is not as severe as plasma electrons are sucked in, thereby modifying the transverse wakefield, but do not cross the axis within the bunch. In {the simulations shown in Fig.~\ref{fig: emittance_evolution}(c)}, $k_b\sigma_{tz}$ increases from 0.09 to 1.4 when $n_b/n_0$ is increased from 1 to 250, explaining the difference in the qualitative evolution of the emittance discussed above. 

These results demonstrate that, for a driver exciting a quasilinear wakefield, the trailing positron bunch rapidly evolves towards an equilibrium with limited emittance growth during this initial evolution, and emittance preservation afterwards. Emittance growth is found to be negligible at the few-percent level for linear and nonlinear positron beam loading when $k_b\sigma_{tz}\ll1$, and moderate for longer bunch lengths and/or higher beam densities with $k_b\sigma_{tz}\gtrsim1$.

\subsection{Longitudinal phase space}
\label{sec:longitudinal}

High-quality beams have not only low transverse emittances, that need to be preserved as discussed in Sec.~\ref{sec:transverse}, but also low longitudinal emittances, that characterize the area occupied by the beam in the longitudinal phase space (LPS). The longitudinal beam quality is most often discussed in terms of total energy spread (including correlated and uncorrelated components) and bunch length, rather than longitudinal emittance.

One important challenge in plasma-based accelerators lies in the large energy chirp that can be imprinted on the accelerated bunch, due to the $\xi$-dependent longitudinal electric field $E_z$, which leads to a large correlated energy spread. This problem has been extensively studied, and can be addressed by either flattening $E_z$ (having a uniform $E_z$ along the bunch)~\cite{Tzoufras2008, Lindstrom2021} or by using a mechanism to dechirp the beam~\cite{Brinkmann2017, Manahan2017, Dopp2018, Darcy2019, FerranPousa2019, Shpakov2019, Wu2019a, Wu2019b, Pompili2021}. While the correlated energy spread can be compensated, the uncorrelated component (or slice energy spread) { is an intrinsic feature of the beam that cannot be removed by dechirping}. Electron acceleration in the blowout regime and in the absence of ion motion has the key advantage that $E_z$ is independent of the transverse coordinate, and therefore the slice energy spread is not degraded during acceleration. This property no longer holds when accounting for ion motion in the blowout regime, and in the case of positron acceleration, the presence of mobile plasma electrons within the accelerated positron bunch induces a transversely non-uniform accelerating field, contributing to an increase on the slice energy spread. Because this degradation of the LPS via the slice energy spread poses a more fundamental limit, we focus the discussion on the uncorrelated energy spread, assuming that the correlated energy spread can be compensated by other means.

\begin{figure}[t!]
\includegraphics[width=0.49\textwidth]{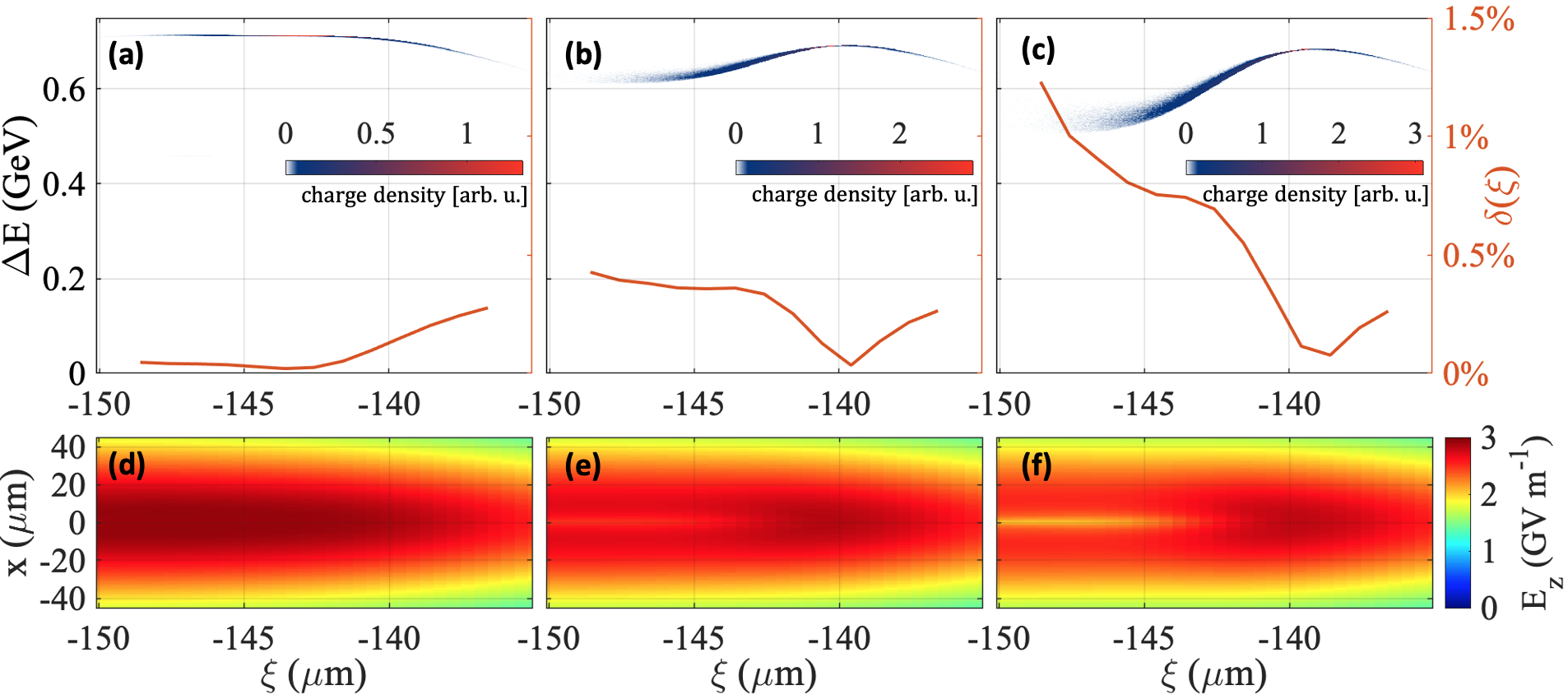}
\caption{\label{fig: fig_4} Simulated LPS of the accelerated trailing positron bunch {[(a)-(c)]} and corresponding longitudinal wakefield map in the vicinity of the positron bunch {[$E_z$, (d)-(f)]} for different initial peak densities: $n_b/n_0=6$ {[(a),~(d)]}, $n_b/n_0=20$ {[(b),~(e)]} and $n_b/n_0=50$ {[(c),~(f)]}. The slice-energy-spread-to-gain ratio $\delta(\xi)$ is represented as orange solid lines on each LPS distribution. $\Delta E$ is the difference between final and initial energy. All plasma, drive and trailing parameters are the same as for simulations in Fig.~\ref{fig: emittance_evolution}(b), keeping the trailing beam size fixed at $\sigma_{tr}=\SI{0.8}{\micro\metre}$ and varying only the trailing charge and $n_b/n_0$.}
\label{fig: fig_4}
\end{figure}

To better assess the growth of slice energy spread in an accelerator stage, we normalize it to the energy gain and define the slice-energy-spread-to-gain ratio $\delta(\xi)$ as
\begin{equation}\label{eq: Uncorrelated energy spread}
    \delta(\xi) = \frac{\Delta E_\mathrm{final}(\xi)}{\langle E_\mathrm{final}\rangle(\xi) -E_\mathrm{init}},
\end{equation}
where $\Delta E_\mathrm{final}(\xi)$ and $\langle E_\mathrm{final}\rangle(\xi)$ are the absolute energy spread and mean energy of slice $\xi$ in the final state, and we have assumed that all particles have an energy $E_\mathrm{init}$ in the initial state. The slice energy spread can vary from head to tail of the bunch and thus generally depends on the slice longitudinal position $\xi$. Assuming a stable wakefield, $\delta(\xi)$ can also be calculated from the knowledge of the longitudinal wakefield $E_z(x,y,\xi)$ and of the bunch density $n_b(x,y,\xi)$:
\begin{equation}
    \delta(\xi) = \frac{1}{\langle E_z\rangle(\xi)}\left[\frac{\int{[E_z(x,y,\xi)-\langle E_z\rangle(\xi)]^2 n_b dxdy}}{\int{n_bdxdy}}\right]^{1/2},
\end{equation}
with 
\begin{equation}
\langle E_z\rangle(\xi)=\frac{\int{E_z(x,y,\xi)n_bdxdy}}{\int{n_bdxdy}}.
\end{equation}

Figure~\ref{fig: fig_4} illustrates how the beam LPS, after \SI{23}{\centi\metre} of acceleration in the quasilinear plasma wakefield excited by an electron drive bunch, and the longitudinal wakefield map of $E_z$, are modified when increasing the peak density of the trailing positron beam, and thus the {energy-transfer} efficiency. It is shown that at high beam density, positron beam loading induces a transversely non-uniform accelerating field $E_z$ {[see Fig.~\ref{fig: fig_4}(d)-(f)]}, especially at the rear of the bunch, leading to an increase of the slice energy spread {[see Fig.~\ref{fig: fig_4}(a)-(c)]}, with $\delta(\xi)$ approaching \SI{1.5}{\%} at the rear of the bunch for $n_b/n_0=50$. The results show that the degradation of the LPS via the uncorrelated energy spread can be severe, considering a \SI{1}{\percent} uncorrelated energy spread as an upper bound for the acceptance of a collider final focus system~\cite{Raimondi2001}. While the drive parameters are kept fixed here as $n_b/n_0$ and $\eta$ are increased, it is possible to optimize the driver, thereby minimize the uncorrelated energy spread at a given level of efficiency $\eta$ (see Sec.~\ref{sec:optimization}). Subsequently, the best performance at each value of $\eta$ can be determined and the trade-off that exists between uncorrelated energy spread and energy-transfer efficiency can be assessed (see Sec.~\ref{sec:tradeoff}).

\section{Blowout regime using a donut-shaped electron beam driver}
\label{sec:donut}

\begin{figure*}[ht!]
\includegraphics[width=1.0\textwidth]{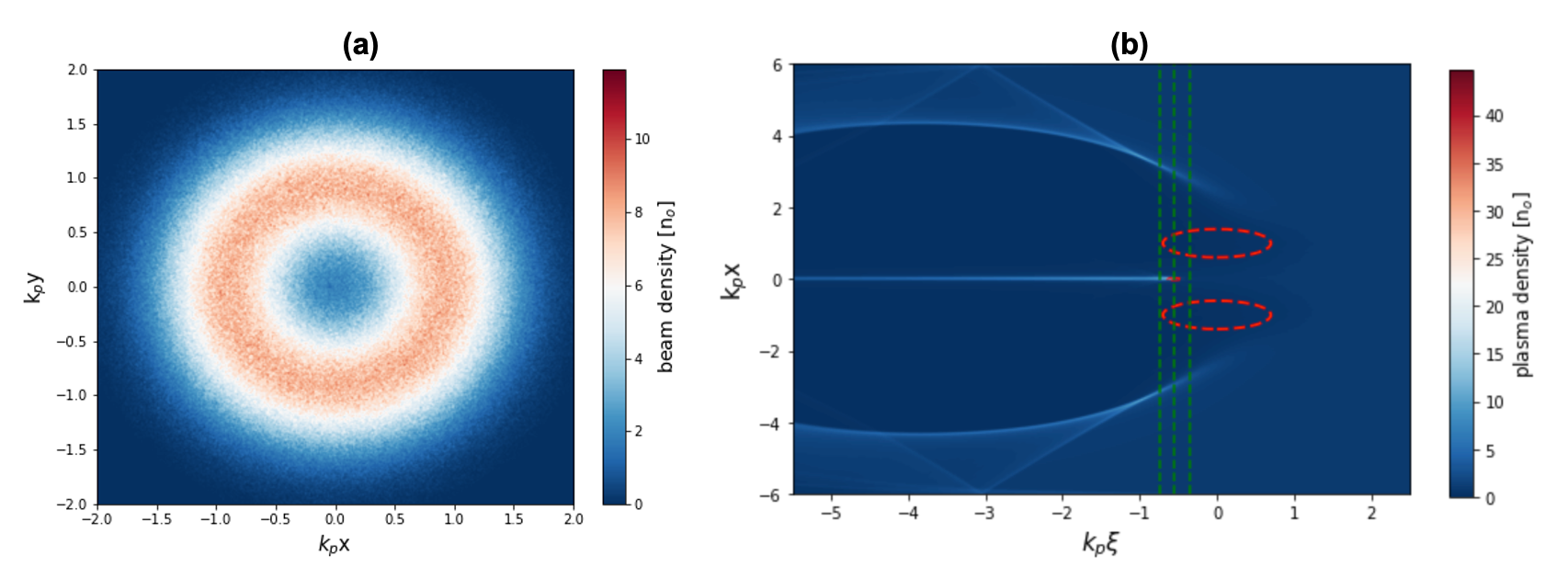}
\caption{\label{DD_collect} Loaded nonlinear plasma wakefield driven by a hollow electron driver. The donut electron driver parameters are $Q_d=\SI{8.6}{nC}$, $k_pr_0 = 1$, $k_p\sigma_r = 0.39$, $k_p\sigma_z=0.7$ and $\xi_0=0$. A Gaussian trailing positron bunch is located at $k_p\xi=-0.55$ with $Q_t=\SI{189}{pC}$, $k_p\sigma_r = 0.035$, and $k_p\sigma_z=0.09$. The plasma density is set to $n_0=\SI{5e16}{\per\cubic\centi\metre}$. (a) Cross section of the donut driver bunch density in the $(x,y)$ plane, (b) map of the plasma electron density in the $(\xi,x)$ plane, with dashed red lines showing the 1$\sigma$ contours of drive and trailing bunch densities and dashed green lines showing the positron central slice and slices at $2\sigma_z$ from the center.}
\end{figure*}

The traditional blowout regime with uniform ion density is known to provide ideal features for electron acceleration: high acceleration gradient, transversely linear focusing force and uniform accelerating field. For positrons, this regime is not favorable unless a uniform electron filament is added inside the blowout cavity, which provides focusing and minimizes the slice energy spread. An electron filament can be present on the propagation axi{s} in the blowout regime when using a donut-shaped electron~\cite{Jain2015, Vieira2016} or laser~\cite{Vieira2014} driver. Here, we focus on the donut electron beam driver, and define the bunch density of the hollow electron driver as follows~\cite{Jain2015}: 
\begin{equation}
    n_b(r,\xi) = n_\mathrm{peak} \exp\left[-\frac{(r-r_0)^2}{2\sigma_r^2}-\frac{(\xi-\xi_0)^2}{2\sigma_z^2} \right],
    \label{eq:donut}
\end{equation}
where the peak of the bunch density, $n_\mathrm{peak}$, is located at $(r_0,\xi_0)$, off the propagation axis, and $r_0$ and $\sigma_r$ represent respectively the radius and thickness of the donut ring. With such hollow structure for the electron driver, plasma electrons with initial radii $r>r_0$ are blown out similarly to the case of a Gaussian-shaped driver, thereby generating a strong wakefield with high accelerating fields, but inner plasma electrons tend to be pushed inward~\cite{Jain2015}{. This separation of plasma-electron trajectories allows for} the formation of a region with an excess of plasma electrons near the propagation axis that can provide focusing for positrons. A cross section of such a hollow electron driver, with $k_pr_0=1$ and $k_p\sigma_r=0.39$, and the generated blowout structure with the trailing positron bunch and the on-axis electron filament are shown in Fig.~\ref{DD_collect}. The simulation was performed using {QuickPIC-OpenSource}. The quality of the plasma electron uniformity near the propagation axis, as experienced by the trailing positron bunch, can be manipulated by driver optimization (see Sec.~\ref{sec:optimization}) and by placing the positron bunch at the appropriate phase of the accelerating field.

\begin{figure}[t!]
\includegraphics[width=0.5\textwidth]{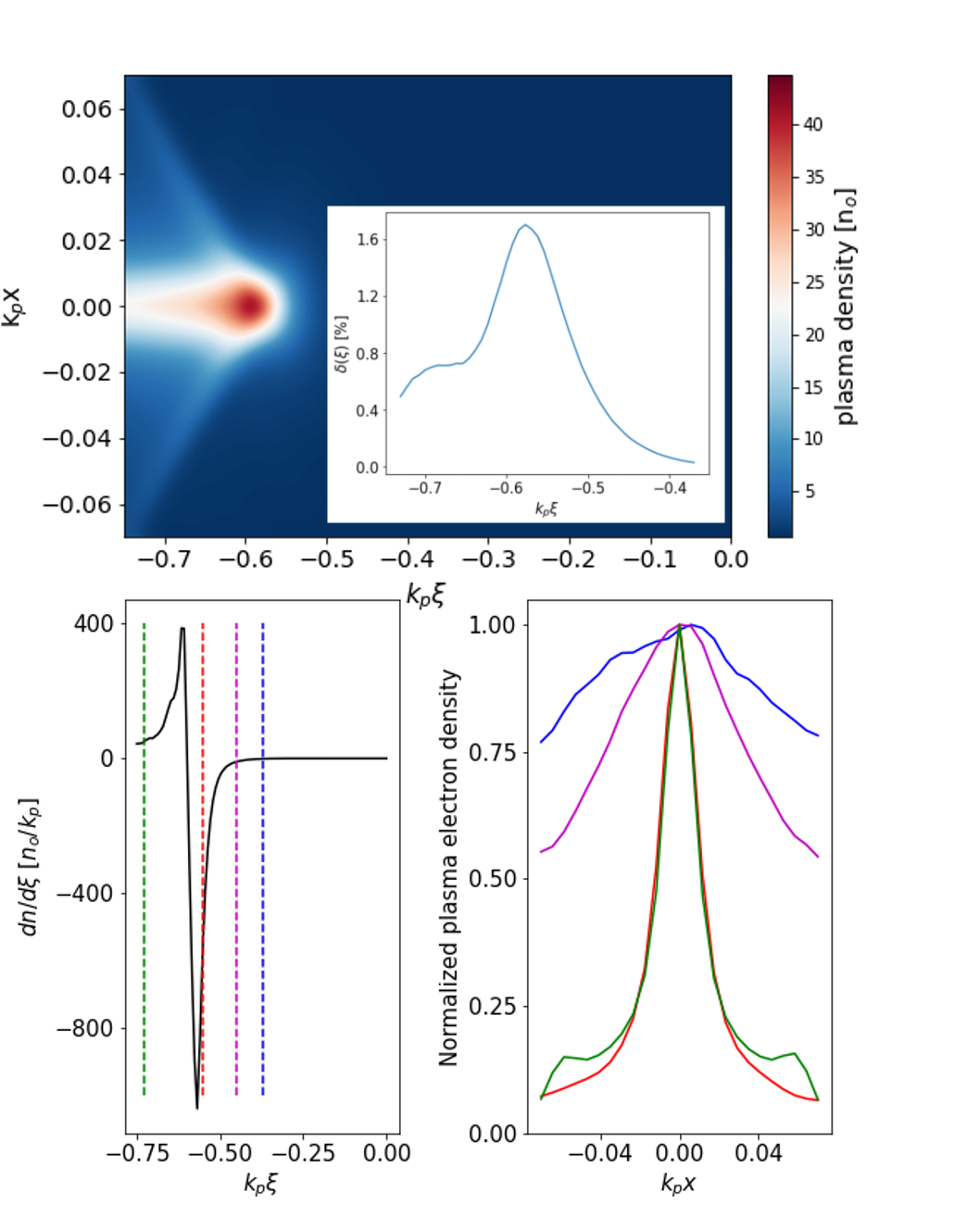}
\caption{\label{DD_beamQua} Plasma electron density in the vicinity of the positron bunch {with the same set of parameters given in Fig. \ref{DD_collect}}: map of $n_p$ in the $(\xi,x)$ plane (up), on-axis value of $dn_p/d\xi$ (bottom left) and transverse line-outs of $n_p$ (bottom right) for different $\xi$ slices (shown as dashed lines in bottom left). The inset of the top plot shows the slice-energy-spread-to-gain ratio $\delta(\xi)$ from $-2\sigma_z$ to $2\sigma_z$ of the positron bunch.}
\end{figure}

Similarly to the discussion in Sec.~\ref{sec:longitudinal}, we would like to find an optimal situation minimizing the slice energy spread across all bunch slices. According to Panofsky-Wenzel theorem~\cite{Panofsky1956}, $\partial_\xi F_r=\partial_r F_z$. {Neglecting ion motion, the contribution of ions and blown-out electrons to the transverse force is a simple $\xi$-independent and transversely linear term $k_p r/2$, and thus the shape of $F_r$ is mainly determined by the plasma electron filament source term. The} best situation can be expected when the plasma electron density is uniform on-axis, and has the same dependence in $r$ off-axis for all longitudinal slices. This can be reasonably achieved over the size of a short and small positron bunch in unloaded or lightly loaded cases. When the wakefield is moderately or heavily loaded, the positron bunch induces a strong response from plasma electrons near the propagation axis, leading to non-uniformity in their density, {both} radially and longitudinally, resulting in sizeable slice energy spread, especially near the bunch center. Figure~\ref{DD_beamQua} shows the electron plasma density in the vicinity of the positron bunch and the corresponding slice-energy-spread-to-gain ratio $\delta(\xi)$, for the same simulation as the one presented in Fig.~\ref{DD_collect}. The plasma density is uniform in $\xi$ and {has} a slow dependence in $r$ at the head, {that is,} when the beam loading is still weak. As a result, the slice energy spread is low at the bunch head. Moving backward along the bunch, the non-uniformity in the plasma density then increases and peaks right after the bunch center, and then drops slightly towards the bunch tail. This behavior in plasma electron uniformity is directly reflected in the slice-energy-spread-to-gain ratio $\delta (\xi)$ of the positron bunch (see inset of upper plot in Fig.~\ref{DD_beamQua}).
\newline \indent In order to maintain the slice-energy-spread-to-gain ratio below the percent level, the trailing-to-drive bunch charge ratio as well as the {energy-transfer} efficiency $\eta$ remain low at the few percent level. Continuing to increase the positron bunch charge results in overloading the wakefield and degrading the beam quality. For the parameters of Figs.~\ref{DD_collect}-\ref{DD_beamQua}, the charge ratio is \SI{2.2}{\percent}, the {energy-transfer} efficiency is \SI{2.9}{\percent}, and the slice energy spread is at the percent level.

\section{Uncorrelated energy spread and energy efficiency}
\label{sec:espread_efficiency}

To determine optimal scenarios for positron acceleration in different schemes and regimes, we focus on two figures of merit: the {energy-transfer} efficiency and the uncorrelated energy spread, as they can be important limits (see Secs.~\ref{sec:linear_wakefields} and~\ref{sec:donut}) for a high-energy collider application. While we have considered so far the slice-energy-spread-to-gain ratio $\delta(\xi)$, which depends on the slice $\xi$, we will use here a single parameter $\delta$ quantifying the longitudinal quality, and defined as 
\begin{equation}
    \delta = \frac{1}{\langle E_z\rangle}\left[\frac{1}{N_b}\int{[E_z(x,y,\xi)-\langle E_z\rangle(\xi)]^2 n_b dxdyd\xi}\right]^{1/2},
\end{equation}
with 
\begin{equation}
\langle E_z\rangle=\frac{1}{N_b}\int{E_z(x,y,\xi)n_bdxdyd\xi}.
\end{equation}
This parameter $\delta$ describes the ratio between the total energy spread of the beam after removal of the chirp induced by $\langle E_z\rangle(\xi)$ and the energy gain, and will be referred to as the uncorrelated-energy-spread-to-gain ratio from now on.

\subsection{Driver optimization}
\label{sec:optimization}

\begin{figure*}[ht!]
    \includegraphics[width=1.0\textwidth]{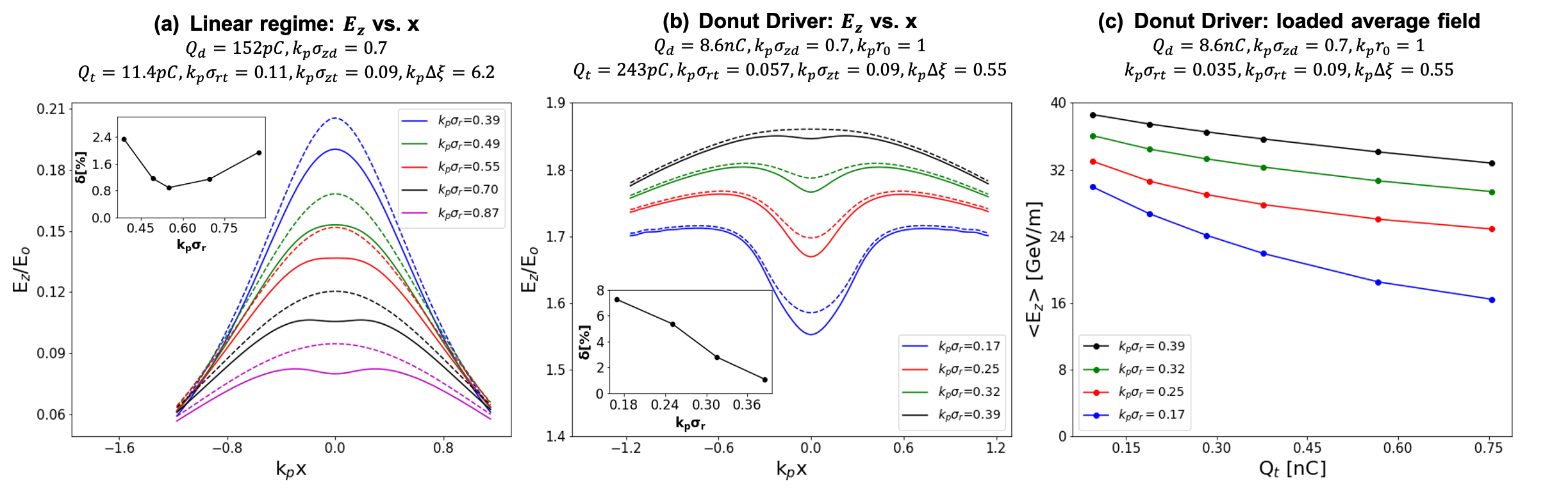}
\caption{The central-slice accelerating field $E_z$ as a function of the transverse coordinate $x$ for the linear regime (a) and for the donut regime (b), for different values of $k_p\sigma_r$ for the driver, and for loaded {(solid lines) and unloaded (dashed lines) cases}. The insets in (a) and (b) show the uncorrelated-energy-spread-to-gain ratio $\delta$. (c) Loaded accelerating field $\langle E_z\rangle_t$ as a function of the positron beam charge $Q_t$ in the donut regime, for drivers with different values of $k_p\sigma_r$. { The plasma density is $n_0=\SI{5e16}{\per\cubic\centi\metre}$ in all simulations}.}
\label{driver_opt}
\end{figure*}

For a given set of positron beam parameters, the driver, and in particular its $\sigma_r$ value, can be optimized in order to minimize $\delta$. Here, we keep the drive beam charge constant so that $\eta$ is not being strongly modified in this optimization. Figure~\ref{driver_opt}(a) highlights the process of drive beam optimization in the linear regime. All simulations were performed using {QuickPIC-OpenSource}. Looking at the transverse dependence of the longitudinal electric field, $E_z$ vs. $x$, and comparing the unloaded (dashed line) to the loaded case (solid line), it is found that beam loading can transversely flatten $E_z$, thereby minimizing the uncorrelated energy spread. This result is valid when the flattening of $E_z$ vs. $x$ is done for the central beam slice, which has the largest effect on the overall beam quality. At a given positron bunch charge, the fields produced by a specific drive bunch is either overloaded, flattened or insufficiently loaded at the central bunch slice. For instance, a large positron charge can quickly overload the fields for drivers with large $\sigma_r$. Therefore, optimizing the drive beam size allows the transverse flattening of $E_z$ which leads to an optimized $\delta$ [see inset of Fig.~\ref{driver_opt}(a)]; this optimization is required for each value of the positron charge. When increasing the positron bunch charge, drivers with smaller $\sigma_r$ are required to maintain a transversely flattened $E_z$ and to keep $\delta$ to a minimum. 

This optimization, however, is ultimately limited if one wants to remain in the linear regime, as the decrease of $\sigma_r$ eventually leads to $n_b>n_0$ for the driver. Here, we will refer to the linear or quasilinear regime when the driver density satisfies $n_b/n_0 \leq 0.5$. Nevertheless, it can be interesting to leave the linear regime and further decrease $\sigma_r$ and increase $n_b$ with a partial blowout forming, as long as a good performance is achieved. This is the case in the moderately nonlinear regime with $n_b/n_0$ going up to 2, above which a complete blowout structure is observed with a degraded longitudinal quality for the accelerated positron bunch. In this moderately nonlinear regime, characterized by a driver with $n_b/n_0\sim 1-2$ and a normalized current $\Lambda = 2I_\mathrm{peak}/I_A=k_p^2\sigma_r^2n_b/n_0<1$ (non-relativistic response of plasma electrons~\cite{Lu2010}, $I_\mathrm{peak}$ being the driver peak current and $I_A\simeq\SI{17}{kA}$ the Alfv\'en current), the drivers are optimized the same way as in the linear regime, yet this optimization only provides small changes in $\delta$. An additional optimization in trailing positron bunch position is performed to ensure that the positron bunch stays in a focusing field.

Figure \ref{driver_opt}(b) shows a similar process for the donut regime. On the one hand, when the donut-shaped electron driver has a large hollow region [i.e. with a small value of $\sigma_r$ in Eq. (\ref{eq:donut})], allowing an excessive amount of plasma electrons to enter in the blowout cavity, it leads to non-uniform plasma electron densities near the propagation axis. In addition, such a driver is more susceptible to beam loading, {as indicated in Fig.~\ref{driver_opt}(c), showing the stronger reduction of the loaded accelerating field with the trailing positron charge, for small values of $\sigma_r$}. In such a regime with the usual plasma electron sheath orbiting around the blowout cavity and with the on-axis plasma electron filament, the positron beam load acts on both plasma species. When decreasing $\sigma_r$ and increasing the number of on-axis plasma electrons, the relative contribution of the latter in the beam loading process becomes more prominent, leading to a stronger reduction of $E_z$ when the positron charge is increased [see Fig.~\ref{driver_opt}(c)]. On the other hand, a relatively uniform on-axis plasma electron density is created by a driver with a large value of $\sigma_r$, allowing sufficient number of plasma electrons to enter the blowout cavity and to provide focusing for the positron bunch, while at the same time keeping non-uniformity in the plasma electron density in the vicinity of the positron bunch at a low level. This results in a much flatter $E_z$ field (transversely) at large $\sigma_r$ in both unloaded and loaded cases [see Fig.~\ref{driver_opt}(b)], and in a lower uncorrelated energy spread [see inset of Fig.~\ref{driver_opt}(b)]. However, the donut thickness $\sigma_r$ cannot be increased indefinitely, as for $k_p\sigma_r\gtrsim0.4$, the plasma electron density on the propagation axis becomes insufficient to compensate for the background ions, resulting in a defocusing field for the positron bunch.

\subsection{Trade-off $\delta - \eta$ for different regimes}
\label{sec:tradeoff}

Generally, the most straightforward way to increase the energy-transfer efficiency $\eta$ is to increase the trailing positron charge, in order to extract more energy from the plasma wakefield. For electron acceleration in the blowout regime, apart from limits associated with ion motion~\cite{Rosenzweig2005, Gholizadeh2010, An2017, Benedetti2017, Mehrling2018} and hosing instability~\cite{Huang2008, Mehrling2017, Lebedev2017}, this increase of charge and efficiency can be done without compromising beam quality, or even with an improved total energy spread by flattening $E_z$ longitudinally and thus minimizing the energy chirp along the accelerated bunch~\cite{Tzoufras2008}. In the case of positron acceleration with plasma electrons providing focusing fields in the vicinity of the trailing bunch, the situation is different and typically higher positron charge and higher $\eta$ comes at the cost of a degraded beam quality and a higher uncorrelated-energy-spread-to-gain ratio $\delta$, because of the strong and fast response of those plasma electrons to the positron load. As a result, there is a clear trade-off between high efficiency and low uncorrelated energy spread, and different regimes can perform differently in this manner.

\begin{figure}[t!]
    \includegraphics[width=0.5\textwidth]{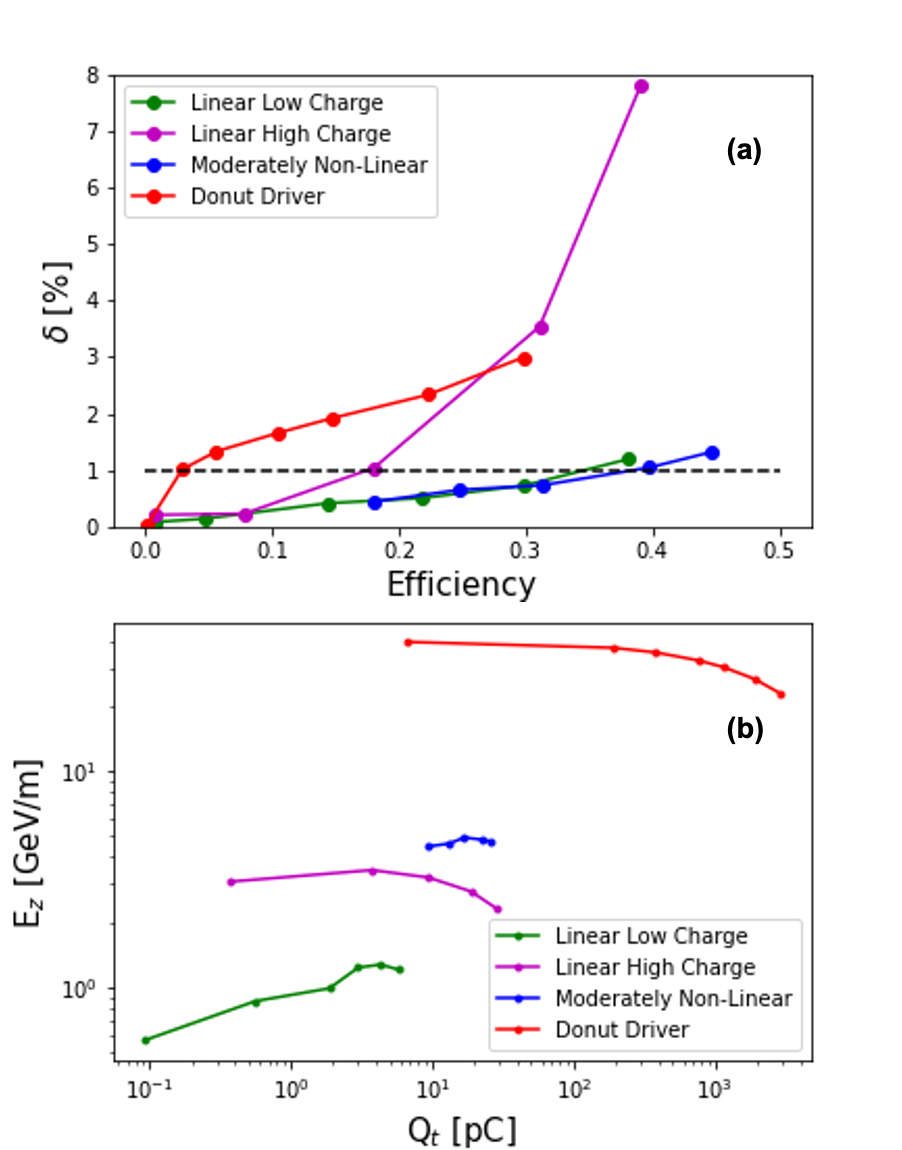}%
\caption{(a) Uncorrelated-energy-spread-to-gain ratio $\delta$ vs. energy-transfer efficiency $\eta$ for different regimes of positron acceleration. The dashed black line depicts the constraint $\delta\lesssim\SI{1}{\percent}$ explained in the text. For each regime, the different data points are obtained by increasing $Q_t$ and optimizing the driver to minimize $\delta$. (b) Loaded accelerating field $E_z$ vs. trailing positron charge $Q_t$ for the same data points as in (a). Beam and plasma parameters or parameter range for these simulation results can be found in Table~\ref{table: final comp param}.}
\label{main_quaeff}
\end{figure}

As higher efficiency is achieved by stronger beam loading, it directly impacts $\delta$. Yet, as discussed in Sec.~\ref{sec:optimization}, as we increase $\eta$ by taking a higher positron charge, the plasma wakefield and its driver need to be re-optimized to provide the lowest uncorrelated energy spread for this higher positron charge and efficiency. This is especially important in the linear regime, as the optimum for $\delta$ is obtained by transverse flattening of $E_z$, which can be controlled by the driver beam size $\sigma_r$.

In order to make relevant comparisons between different schemes, we keep similar parameters for the trailing positron bunch, namely a beam size $\sigma_r$ of around $\SI{1}{\micro\metre}$, and the same bunch length, $\sigma_z=\SI{2.14}{\micro\metre}$. The drive bunch length is also kept fixed at $\sigma_z=\SI{16.7}{\micro\metre}$, as well as the plasma density at $n_0=\SI{5e16}{\per\cubic\centi\metre}$. All parameters or parameter range used in the simulations are summarized in Table \ref{table: final comp param}. The determination of $\eta$ and $\delta$ with single-step {QuickPIC-OpenSource} simulations does not depend on the positron beam energy and emittance, but given the positron beam sizes considered here, quasi-matching (see Sec.~\ref{sec:transverse}) would typically require \si{\micro\metre}-scale normalized emittance for a \SI{1}{GeV} beam. Simulating scenarios relevant to lower emittances and higher energies require much smaller beam sizes and thus extremely high resolution, which is beyond the scope of this paper.

The result of this comparison is presented in Fig.~\ref{main_quaeff}(a), where for each regime increasing values of $\eta$ are obtained by increasing the trailing positron charge $Q_t$, and the optimization described in~\ref{sec:optimization} is performed to minimize $\delta$ for each individual data point. Figure~\ref{main_quaeff}(a) thus provides a direct representation of the trade-off between $\delta$ and $\eta$ for different regimes.

In the linear regime with a low drive beam charge of \SI{38}{pC} (referred to as ``Linear Low Charge'' in Fig.~\ref{main_quaeff}), one finds the trailing positron charge and energy-transfer efficiency can be increased to about \SI{5}{pC} and \SI{30}{\percent}, while keeping $\delta$ below \SI{1}{\percent}. Here, we will consider this \SI{1}{\percent} limit in $\delta$, shown as a dashed line in Fig.~\ref{main_quaeff}(a) as an upper bound for an acceptable uncorrelated energy spread in view of the requirements of a final focus system~\cite{Raimondi2001}. Because of the small drive charge, it's possible to keep decreasing the driver beam size $\sigma_r$ as $Q_t$ and $\eta$ are increased, and to continue to excite a linear wakefield with $n_b/n_0\leq0.5$ for the driver. However, with such low drive charge, the loaded accelerating field only slightly exceeds \SI{1}{\giga\volt\per\metre} [see Fig.~\ref{main_quaeff}(b)], and the accelerated positron charge is only \SI{5}{pC}. 

\begin{table*}[t!]
\begin{center}{}
\resizebox{0.9\textwidth}{!}{
\begin{tabular}{ p{0.1\textwidth}|p{0.1\textwidth}|p{0.1\textwidth}|p{0.1\textwidth}|p{0.1\textwidth}|p{0.1\textwidth}|p{0.1\textwidth}|p{0.1\textwidth}|p{0.1\textwidth} }
    & \multicolumn{2}{c|}{\textbf{Linear Low Charge}}
            & \multicolumn{2}{c|}{\textbf{Linear High Charge}}
                    & \multicolumn{2}{c|}{\textbf{Moderately Non-Linear}}
                            & \multicolumn{2}{c}{\textbf{Donut Driver}} \\
    \hline
     & Driver & Trailing & Driver & Trailing & Driver & Trailing & Driver & Trailing \\
     \hline
     $\sigma_{r}\:(\si{\micro\metre})$ &6.09--19.27 & 1.19 & 12.19--14.56 & 1.19 & 6.28--8.22 & 1.19 & 9.4 &0.85 \\
     $\sigma_{z}\:(\si{\micro\metre})$ &16.7 &2.14 &16.7 &2.14 &16.7 &2.14 &16.7& 2.14 \\
     $n_b/n_0$ &0.05--0.5&0.25--15.5&0.35--0.5&1--75&1.1--1.88&25--70&2.97&35--15000 \\
     $k_p\xi$ &0&-6.2&0&-6.2&0&-6.25 -- -5.90&0&-0.55 \\
\end{tabular}
}
\caption{Beam parameters or parameter range for the simulation results presented in Fig.~\ref{main_quaeff}. The plasma density is $n_0=\SI{5e16}{\per\cubic\centi\metre}$ for all simulations, and $k_pr_0=1$ for ``Donut Driver'' simulations. In all regimes, for a trailing positron beam with an initial energy $E=\SI{1}{GeV}$, quasi-matched conditions generally correspond to \si{\micro\metre}-scale normalized emittances.}
\label{table: final comp param}
\end{center}
\end{table*}

One can naturally seek to accelerate higher trailing positron charges at higher fields by increasing the drive beam charge. Considering the linear regime again but with a higher drive beam charge of \SI{152}{pC} (referred to as ``Linear High Charge'' in Fig.~\ref{main_quaeff}), and repeating the same optimization process, one finds that the limit of $n_b/n_0\leq0.5$ for the driver to continue to excite linear wakefields prevents any further optimization or decrease of the drive beam size $\sigma_r$ beyond the first two ``Linear High Charge'' data points in Fig.~\ref{main_quaeff}. Because of this lack of optimization, $\delta$ quickly increases for the following data points with higher $Q_t$ and $\eta$. Using the same constraint $\delta\lesssim\SI{1}{\percent}$ as before, $Q_t$ and $\eta$ are limited to about \SI{10}{pC} and \SI{20}{\percent}, while the loaded accelerating field $E_z$ reaches \SI{3}{\giga\volt\per\metre}.

In fact, we can continue to decrease the drive beam size $\sigma_r$ despite leaving the linear regime, as long as the acceleration performance is satisfactory, which is found to be the case in the moderately nonlinear regime characterized by $n_b/n_0\sim 1-2$ and $\Lambda<1$, and introduced in Sec.~\ref{sec:optimization}. In Fig.~\ref{main_quaeff}, the data points for the moderately nonlinear regime share the same drive charge of \SI{152}{pC} as the ``Linear High Charge'' case, but the smaller value of $\sigma_r$ being used for the driver makes it possible to considerably improve the transverse uniformity of $E_z$, resulting in lower $\delta$. Figure~\ref{main_quaeff} shows that in the moderately nonlinear regime we can achieve $\delta\lesssim\SI{1}{\percent}$ with trailing positron charge and energy-transfer efficiency of up to \SI{25}{pC} and \SI{40}{\percent}, and an accelerating field of $E_z\simeq\SI{5}{\giga\volt\per\metre}$.

Finally, if one aims for even higher accelerating field and higher positron charge, Fig.~\ref{main_quaeff} shows that the nonlinear blowout regime with a donut-shaped electron driver is the best suited, at the cost of a degraded trade-off between $\delta$ and $\eta$. This regime is indeed compatible with much higher drive charge ($Q_d=\SI{8.6}{nC}$ for the ``Donut Driver'' data points in Fig.~\ref{main_quaeff}), and thus higher $E_z$ and $Q_t$, typically one to two orders of magnitude higher than in the previous cases. However, above \SI{5}{\percent} energy-transfer efficiency, $\delta$ degrades beyond \SI{1}{\percent}, so the donut regime can be used for high fields and high trailing charges with a compromise on either an energy-transfer efficiency limited to the few-percent range, or on an uncorrelated energy spread exceeding the percent level.

\section{Conclusion}
\label{sec:conclu}

Our results show the importance of beam loading in plasma-based positron acceleration, whose properties differ significantly from beam loading for electron acceleration in the blowout regime. Indeed, when plasma electrons are present in the vicinity of the accelerated positron bunch to provide focusing fields, positron beam loading can then alter the focusing properties of the plasma wakefield as well as the transverse uniformity of the accelerating field, because of the strong and fast response of those plasma electrons. Yet, beam loading is a prerequisite {for} good energy-transfer efficiency, which is highly desirable for a high energy physics application. The results presented in the previous section show that one needs to make a compromise between the uncorrelated energy spread described by the parameter $\delta$ and the energy-transfer efficiency $\eta$.

In the linear regime, we have found that while generally the positron bunch quickly evolves towards an equilibrium with limited emittance growth, the uncorrelated energy spread can set an important limit. To maximize the energy-transfer efficiency in the linear regime, both the electron driver and the trailing positron bunch should have small beam sizes, $k_p\sigma_r < 1$, to avoid leaving energy in the plasma wave by poor matching of the transverse size of the drive and trailing plasma wakefields. Interestingly, driving a linear plasma wakefield with a Gaussian-shaped electron driver with $n_b/n_0<1$ and extracting its energy with nonlinear beam loading by a trailing positron bunch with $n_b/n_0>1$ is fully relevant and provides good performance with $\eta$ going up to \SI{30}{\percent} while keeping $\delta\lesssim\SI{1}{\percent}$. The limited positron charge and accelerating field of the linear regime can be increased in the moderately nonlinear regime, with $\eta$ going up to \SI{40}{\percent} for $\delta\lesssim\SI{1}{\percent}$. The nonlinear blowout regime driven by a donut-shaped electron driver is found to allow acceleration of much higher positron charge at much higher accelerating gradients, but the energy-transfer efficiency needs to be kept at the few-percent level to fulfill $\delta\lesssim\SI{1}{\percent}$.

Further research will aim at extending the results to lower emittances, asymmetric beams and higher energies that are relevant in the route towards a plasma-based collider design, and to provide systematic comparisons of existing positron acceleration regimes. It is important to note that a $e^+e^-$ collider does not necessarily require the same performance for electrons and positrons, and in particular for a plasma-based collider, the requirements for positrons could be not as stringent as those for electrons, and thus may be somewhat {lowered} in comparison to the parameters of linear collider designs based on RF accelerators~\cite{ilc,clic}. Given the challenges of positron acceleration in plasmas, {having asymmetrical requirements for electrons and positrons} could be crucially important towards realistic designs of plasma-based colliders \cite{Chen:2020cxz}.

\begin{acknowledgments}
The work at LOA was supported by the European Research Council (ERC) under the European Union’s Horizon 2020 research and innovation programme (Miniature beam-driven Plasma Accelerators project, Grant Agreement No. 715807). SLAC was supported by U.S. DOE Contract DE-AC02-76SF00515. The work at University of Oslo was supported by the Research Council of Norway, Grant No. 313770. Numerical simulations were performed using HPC resources from GENCI-TGCC (Grants No. 2020-A0080510786 and No. 2020-A0090510062) with the IRENE supercomputer, and using the Open Source version~\cite{QuickPIC-OpenSource} of the QuickPIC code~\cite{Huang2006, An2013}, a 3D parallel (MPI \& OpenMP Hybrid) Quasi-Static Particle-In-Cell code.
\end{acknowledgments}

\appendix

\section{energy-transfer efficiency in three-dimensional linear plasma wakefields}
\label{Appendix_eta3D}

In a two-bunch acceleration scheme, energy is transferred from the driver to the plasma wave, and the trailing bunch takes the energy from the plasma wave excited by the driver. Apart from energy-transfer efficiency calculated using the particle point of view shown in Eq.~\eqref{eq: eta definition}, the efficiency can also be obtained by calculating the energy in the plasma waves generated by the drive and trailing bunches:
\begin{equation}\label{Appendix eq: efficiency, before, after}
    \eta = 1-\frac{\int E_{z0,\mathrm{tot}}^2d^2\mathbf{r}_\perp+\int E_{r0,\mathrm{tot}}^2d^2\mathbf{r}_\perp}{\int E_{z0,d}^2d^2\mathbf{r}_\perp + \int E_{r0,d}^2d^2\mathbf{r}_\perp},
\end{equation}
where $E_{z0,d}$ and $E_{r0,d}$ are the amplitudes of the $z$ and $r$ components of the electric field of the drive beam plasma wave, $E_{z0,\mathrm{tot}}$ and $E_{r0,\mathrm{tot}}$ are the amplitudes for the total plasma wave from both beams, and the integral is performed over the transverse coordinates. Importantly, these amplitudes need to be evaluated in the wake of the beams.

In the linear regime, the longitudinal and radial components of the electric field for the plasma wave excited by a particle beam with cylindrical symmetry can be written as~\cite{Keinigs1987}:
\begin{align}
    \nonumber
    E_z(r,\xi) = &\frac{qk_p^2}{\epsilon_0}\int_0^{+\infty} r^\prime dr^\prime  K_0(k_pr_>)I_0(k_pr_<)\\
    &\times\int_\xi^{+\infty} d\xi^\prime n_b(r^\prime,\xi^\prime) \cos{k_p(\xi-\xi^\prime)},\\
    \nonumber
    E_r(r,\xi) = &-\frac{qk_p}{\epsilon_0}\int_0^{+\infty} r^\prime dr^\prime K_1(k_pr_>)I_1(k_pr_<)\\
    &\times\int_\xi^{+\infty} d\xi^\prime \frac{\partial n_b(r^\prime,\xi^\prime)}{\partial r^\prime} \sin{k_p(\xi-\xi^\prime)},
\end{align}
where $n_b(r,\xi)$ is the bunch density, $q$ is the particle charge, $r_<$ is the smaller of $r$ and $r^\prime$ and $r_>$ the larger. $I_n$ and $K_n$ are the $n^\mathrm{th}$ order modified Bessel functions of the first and second kind respectively. 

\subsection{Efficiency for separable beams}

When the beam density is mathematically separable in the coordinates $r$ and $\xi$, that is $n_b(r,\xi) = N \mathcal{R}(r) \mathcal{Z}(\xi)$ with $N$ the number of particles in the beam, $E_z(r,\xi)$ and $E_r(r,\xi)$ can also be written in separable forms. We use the following convention for $\mathcal{R}$ and $\mathcal{Z}$:
\begin{align}
    \nonumber &\int_0^{+\infty} 2\pi r dr \mathcal{R}(r) = 1,\\
    &\int_{-\infty}^{+\infty} d\xi\mathcal{Z}(\xi) = 1,
\end{align}
for which the condition $\int n_b d^3\mathbf{r} = N$ is satisfied. In this case, $E_z(r,\xi)$ and $E_r(r,\xi)$ can be written as:
\begin{align}
    \nonumber
    &E_z(r,\xi) = \frac{Nqk_p^2}{\epsilon_0} \mathbb{L}_z(\xi)\mathbb{T}_z(r),\\
    &E_r(r,\xi) =  -\frac{Nqk_p^2}{\epsilon_0} \mathbb{L}_r(\xi)\mathbb{T}_r(r),
    \label{Appendix eq: Electric field in three parts}
\end{align}
with
\begin{align}
    \nonumber
    &\mathbb{L}_z(\xi) = \int_\xi^\infty d\xi^\prime \mathcal{Z}(\xi^\prime)\cos{k_p(\xi-\xi')},\\    \nonumber
    &\mathbb{L}_r(\xi) = \int_\xi^\infty d\xi^\prime \mathcal{Z}(\xi^\prime)\sin{k_p(\xi-\xi')},\\
    \nonumber
    &\mathbb{T}_z(r) = \int_0^\infty r^\prime dr^\prime K_0(k_pr_>)I_0(k_pr_<) \mathcal{R}(r'),\\
    &\mathbb{T}_r(r) = \frac{1}{k_p}\int_0^\infty r^\prime dr^\prime K_1(k_pr_>)I_1(k_pr_<) \frac{\partial \mathcal{R}(r^\prime)}{\partial r^\prime},
    \label{Appendix eq: def LT}
\end{align}
where $\mathbb{L}_{z,r}$ describes the longitudinal shape and only depend on $\xi$, and $\mathbb{T}_{z,r}$ describes the transverse shape and only depend on $r$. $\mathbb{L}_{z,r}$ and $\mathbb{T}_{z,r}$ are all dimensionless functions.

In Eq.~\eqref{Appendix eq: efficiency, before, after}, we need the amplitude of the fields in the wake of the beams, which can be mathematically evaluated in the limit $\xi\rightarrow -\infty$ when beams have finite lengths. In this limit, both $\mathbb{L}_z$ and $\mathbb{L}_r$ are sinusoidal functions of $\xi$, which are \SI{90}{degrees} out of phase with each other, but share the same maximum, which we will simply denote as $\mathbb{L}$ in the following. We have for this longitudinal factor $\mathbb{L}$:
\begin{align}
    \nonumber \mathbb{L} &= \max_\phi \int_{-\infty}^{+\infty} d\xi^\prime \mathcal{Z}(\xi^\prime)\cos{k_p(\phi-\xi')}\\ 
    \nonumber &=\max_\phi \left[\mathrm{Re} \int_{-\infty}^{+\infty} d\xi^\prime \mathcal{Z}(\xi^\prime)e^{ik_p(\phi-\xi')} \right]\\
    \nonumber &=\max_\phi \left[\mathrm{Re}\; e^{ik_p\phi} \int_{-\infty}^{+\infty} d\xi^\prime \mathcal{Z}(\xi^\prime)e^{-ik_p\xi'} \right]\\
    &= |\tilde{\mathcal{Z}}(k=k_p)|,
\end{align}
where $\tilde{\mathcal{Z}}$ is the Fourier transform of $\mathcal{Z}$:
\begin{align}
    \nonumber \tilde{\mathcal{Z}}(k) = \int_{-\infty}^{+\infty} dx \mathcal{Z}(x)e^{-ikx}.
\end{align}
In particular, for short bunches with $k_p\sigma_z\ll1$, $\mathbb{L}\simeq1$.

Using Eqs.~\eqref{Appendix eq: efficiency, before, after} and \eqref{Appendix eq: Electric field in three parts}-\eqref{Appendix eq: def LT}, and under the same assumptions as in Eq.~\eqref{eq: 1D model} for the drive-trailing bunch separation $\Delta \xi$ and particle charges $q_d=\pm q_t$, the efficiency for separable beams in the 3D linear regime can be calculated, and reads:
\begin{align}
    \nonumber \eta = &2\frac{N_t\mathbb{L}_t}{N_d\mathbb{L}_d} \frac{\int rdr\left(\mathbb{T}_{z,d}\mathbb{T}_{z,t}+\mathbb{T}_{r,d}\mathbb{T}_{r,t}\right)}{\int rdr\left(\mathbb{T}_{z,d}^2+\mathbb{T}_{r,d}^2\right)}\\ 
    & -\left(\frac{N_t\mathbb{L}_t}{N_d\mathbb{L}_d}\right)^2 \frac{\int rdr \left(\mathbb{T}_{z,t}^2+\mathbb{T}_{r,t}^2\right)}{\int r dr \left(\mathbb{T}_{z,d}^2+\mathbb{T}_{r,d}^2\right)},
    \label{eq: theoretical model for efficiency general}
\end{align}
where the subscripts $d$ and $t$ specify driver and trailing for each quantity. 

\subsection{Efficiency for 3D Gaussian beams}

The beam density for a 3D Gaussian beam writes as:
\begin{equation}\label{Appendix eq: 3D Gaussian beams}
    n_b (r,\xi) = \frac{N}{\sqrt{2\pi}^3\sigma_r^2\sigma_z}\exp{\left(-\frac{r^2}{2\sigma_r^2}\right)}\exp{\left(-\frac{\xi^2}{2\sigma_z^2}\right)}.
\end{equation}
It is a particular case of a separable beam with:
\begin{align}
    \nonumber \mathcal{R}(r) = \frac{1}{2\pi\sigma_r^2}\exp{\left(-\frac{r^2}{2\sigma_r^2}\right)},\\
    \mathcal{Z}(\xi) = \frac{1}{\sqrt{2\pi}\sigma_z}\exp{\left(-\frac{\xi^2}{2\sigma_z^2}\right)}.
\end{align}
The functions $\mathbb{T}_z$ and $\mathbb{T}_r$ are given by Eq.~\eqref{Appendix eq: def LT}, the longitudinal factor $\mathbb{L}$ simplifies to:
\begin{align}
    \mathbb{L} = \exp \left(-\frac{k_p^2\sigma_z^2}{2}\right),
\end{align}
and the energy-transfer efficiency can be calculated using Eq.~\eqref{eq: theoretical model for efficiency general}.

When the beam transverse size is far larger than the plasma skin depth, $\sigma_r\gg 1/k_p$, the function $\mathbb{T}_z$ approximately takes a Gaussian shape $\mathbb{T}_z\sim\exp{(-r^2/2\sigma_r^2)}/\sigma_r^2$, and $\mathbb{T}_r\ll\mathbb{T}_z$. When both drive and trailing bunches satisfy $k_p\sigma_r\gg1$, the expression for $\eta$ can be simplified using the asymptotic approximation for the transverse shape and reads:
\begin{align}
    \eta = \frac{N_t\mathbb{L}_t}{N_d\mathbb{L}_d}\frac{\sigma_{dr}^2}{\sigma_{tr}^2} \left[\frac{4}{1+\frac{\sigma_{dr}^2}{\sigma_{tr}^2}}-\frac{N_t\mathbb{L}_t}{N_d\mathbb{L}_d}\right].
    \label{eq: complete energy-transfer efficiency theory for large beam}
\end{align}
For either short bunches ($k_p\sigma_z\ll1$ and $\mathbb{L}\simeq1$) or for equal bunch length ($\sigma_{dz}=\sigma_{tz}$ and thus $\mathbb{L}_d=\mathbb{L}_t$), we finally obtain Eq.~(\ref{eq: efficiency for same dw bunch length}):
\begin{equation}
    \eta = \frac{N_{t}}{N_{d}}\frac{\sigma_{dr}^2}{\sigma_{tr}^2}\left[\frac{4}{1+\frac{\sigma_{dr}^2}{\sigma_{tr}^2}}-\frac{N_t}{N_d}\right].
\end{equation}
In this case, $\eta$ is maximum for the following values:
\begin{align}
\label{eq:eta_max}
&\eta_\mathrm{max} =\frac{\sigma_{dr}^2}{\sigma_{tr}^2}\left(\frac{2}{1+\frac{\sigma_{dr}^2}{\sigma_{tr}^2}}\right)^2,\\
\label{eq:x_max}
&\left(\frac{N_t}{N_d}\right)_\mathrm{max} =\frac{2}{1+\frac{\sigma_{dr}^2}{\sigma_{tr}^2}}.
\end{align}


%

\end{document}